\documentclass[journal]{IEEEtran}

\usepackage{cite}
\usepackage{amsmath,amssymb,amsfonts}
\usepackage{algorithmic}
\usepackage{graphicx,subfigure}
\usepackage{array}
\usepackage{tabularx}
\usepackage{multirow}
\usepackage{breqn}
\graphicspath{{images/}}
\usepackage{textcomp}
\usepackage{xcolor}
\usepackage{epstopdf}
\usepackage[switch]{lineno}

\begin{document}
\title{NeighCNN: A CNN based SAR Speckle Reduction using Feature preserving Loss Function}
\author{{Praveen Ravirathinam, Darshan Agrawal, J. Jennifer Ranjani \\}

Department of Computer Science and Information Systems, Birla Institute of Technology and Science, Pilani, Rajasthan - 333 031, India.\\ e-mail: {praveen.ravirathinam@gmail.com, darshanagrawal0103@gmail.com, j.jenniferranjani@yahoo.co.in}}

\maketitle

\begin{abstract}
    Coherent imaging systems like synthetic aperture radar are susceptible to multiplicative noise that makes applications like automatic target recognition challenging. In this paper, NeighCNN, a deep learning-based speckle reduction algorithm that handles multiplicative noise with relatively simple convolutional neural network architecture, is proposed. We have designed a loss function which is an unique combination of weighted sum of Euclidean, neighbourhood, and perceptual loss for training the deep network.  Euclidean and neighbourhood losses take pixel-level information into account, whereas perceptual loss considers high-level semantic features between two images. Various synthetic, as well as real SAR images, are used for testing the NeighCNN architecture, and the results verify the noise removal and edge preservation abilities of the proposed architecture. Performance metrics like peak-signal-to-noise ratio, structural similarity index, and universal image quality index are used for evaluating the efficiency of the proposed architecture on synthetic images.

\end{abstract}

\begin{IEEEkeywords}
Despeckling, SAR, CNN, deep neural networks, perceptual loss, neighborhood loss 
\end{IEEEkeywords}

\section{Introduction}
Synthetic aperture radar (SAR) is a coherent and active imaging system that yields high-resolution images all-day independent of the weather conditions. It is suitable for applications such as monitoring crop cultivation, land survey, encroachment tracking, migration, target recognition in the military, disaster management, etc. Irrespective of the high resolution, SAR images suffer from multiplicative speckle noise due to destructive interference of the radio waves transmitted to the target surface. The severity of the speckle limits the suitability of a SAR image as it can skew the region boundaries leading to an error in segmentation, detection, and recognition. Despeckling SAR images has been an active topic of research over the past decade. The multiplicative noise in an $L-$look SAR image is assumed to follow Gamma distribution \cite{gamma}, given by

\begin{equation}
p_N(N) = \frac{L^LN^{L-1}e^{-LN}}{\Gamma(L)}
\label{gamma}
\end{equation}

where $N$ is the multiplicative noise of the SAR image in intensity format, $L$ is the number of looks. The mean and variance of the Gamma distribution are $1$ and $\frac{1}{L}$ respectively.

Earlier despeckling methods focus on image processing or pattern recognition based techniques and are proven to be effective \cite{jranjani11} - \cite{ferraioli}. A category of speckle removal algorithms takes logarithmic transformation to convert multiplicative speckle into an additive noise, but it adds a bias to the despeckled image  \cite{argenti}. Often these algorithms fail to preserve edge and texture details as they process the information locally. Recently, efficient and accurate despeckling algorithms utilizing deep learning have been designed \cite{idcnn} - \cite{liu}. However, a detailed ablation study on the optimal no. of convolution layers and the loss functions for SAR images were unexplored to the best of our knowledge. 

In this work, we have proposed an end to end deep learning architecture using Convolutional Neural Network (CNN) for despeckling SAR images. Earlier CNN based architectures like IDCNN\cite{idcnn}, and SARCNN\cite{sarcnn} use logarithmic transformation or division operation to handle the multiplicative nature of the speckle noise. In the proposed work, we have utilized a simple subtraction layer to remove the residual speckle from the SAR images. Also, a unique loss function that is a combination of neighbourhood, weighted perceptual, and Euclidean loss, is proposed. In addition to the terms in the total variation loss, the proposed neighbourhood loss includes the difference across the diagonals as well, thus accounting for additional coherence. The modified perceptual loss introduces a weighting factor that gives more consideration to the later layers of the feature extraction network. We have designed the proposed three-part loss function to achieve increased accuracy together with an optimal smoothing and edge preservation capabilities. Extensive experiments prove the significance of the three-part loss function. The organization of the paper is as follows: Section II describes the proposed NeighCNN architecture and the optimized loss function. Section III mentions the implementation details and the dataset. Section IV and V evaluates the performance of the proposed algorithm and summarizes the proposed work, respectively.

\section{NeighCNN Speckle Reduction Algorithm}
\subsection{Network Architecture}
Mostly utilized denoising architectures have a series of convolutional layers with batch normalization and ReLU activation function \cite{thakur}\cite{tian}. Variation in these architectures might be the number of convolutional layers or additional residual layers. There are two basic approaches to address the despeckling problem in SAR images: either to process the multiplicative noise directly or to convert the multiplicative noise into an additive one using logarithmic transformation.  

In the proposed approach, we have incorporated a commonly utilized representation \cite{map} to represent the multiplicative noise as
\begin{equation}
Y = X * N = X + X(N-1) = X + X.N' = X + \eta
\label{sar}
\end{equation}

here $X$, $Y$, and $N$ denote the noise-free reflectivity, the noisy signal observed, and the stationary uncorrelated random process with unit mean that is independent of $X$, respectively. Pixel indices in eqn. (\ref{sar}) are omitted for simplicity. $\eta$ here is a signal-dependent, additive noise term with zero mean. CNNs have recently become an efficient tool for extracting underlying and internal features without requiring complex constraints. The training data and corresponding labels can learn the non-linear characteristics of the SAR images from the parameters of the deep network \cite{sardrn}.

\begin{figure*}[htbp!]
\centering
    \includegraphics[scale = 0.5]{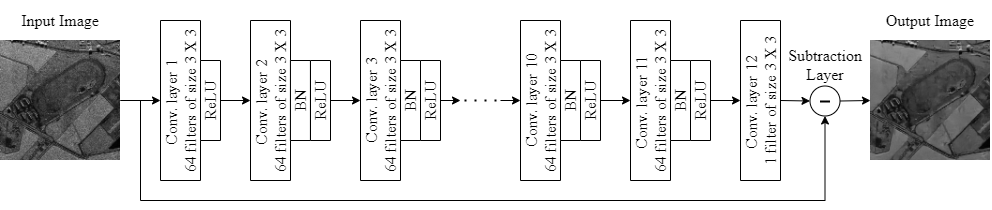}
    \caption{NeighCNN architecture}
    \label{fig:architecture}
\end{figure*}

The proposed architecture shown in fig. \ref{fig:architecture} comprises of 12 convolutional layers. Except for the last, all the other layers use 64 filters. The most commonly used activations functions are sigmoid and tanh. However, the sigmoid function moves all the weights towards the positive or negative direction during gradient descent, and the speed of the tanh function is relatively slow. In \cite{relu}, a rectified linear unit (ReLU) to solve problems that arise from using sigmoid and tanh activation function, is proposed. Also, ReLU ensures that the despeckled image maps the highly non-linear relationship between the speckled and its ground truth. Batch normalization (BN) is incorporated in the intermediate layers to attenuate the covariate shift. BN, together with ReLU, regularizes the data by accelerating the learning and reducing drastically the number of epochs required for training.

The deep networks can avoid vanishing gradient problem by using skip connections between the input and the final layer. Skip connections are used to subtract the information in these two layers \cite{sarcnn} \cite{sardrn} or perform division between them followed by a non-linear $tanh$ function \cite{idcnn}. Residual networks make full use of the skip connections by transferring information between two disconnected layers without attenuation \cite{residual}. Thus, residual learning can explore the ability of the deep network to express non-linear relations.  

\subsection{Loss function}
    The loss function is a vital factor defining a denoising architecture.  The deep network tries to optimize the loss function computed between the actual and the predicted image tensors. In other words, the loss function gives a quantitative metric to represent the loss of data/quality between the input and the predicted images. Loss functions like mean square error, Euclidean loss in $L2-$norm aim to minimize the difference in pixels between the actual and predicted image, but they fail to produce smooth image regions. Some loss functions are computationally complex as they include logarithmic or trigonometric operations.

The proposed three-part loss function is a combination of Euclidean, perceptual loss, and neighbourhood loss. The loss function used in the proposed architecture is given by,

\begin{equation}
         L_T = L_{Eu} + \alpha_NL_{Per} +\beta_NL_N
    \label{totalloss}
    \end{equation}
    
    The total loss, \(L_T\), is defined as a combination of Euclidean loss, \(L_{Eu}\), perceptual loss, \(L_{Per}\), and neighbourhood loss, \(L_N\). Euclidean loss impacts the accuracy as it finds the pixel-wise difference between the two images. But it identifies two images as dissimilar even when there is minimal variation between them. The perceptual loss can rectify this issue as it captures the semantic similarity using high-level features. And neighbourhood loss exploits the coherence between all the adjacent pixels, and it controls the overall degree of smoothness in the despeckled image.  The tuning parameters  \(\alpha_N\), \(\beta_N\) controls the significance of the perceptual and neighbourhood loss, and they normalize the scale difference between the two components.  The values of both \(\alpha_N\) and \(\beta_N\) should always be less than \(1\) to achieve increased accuracy and to avoid over-smoothing. 
    
The Euclidean loss in eqn. (\ref{totalloss}) is given by, 
    \begin{equation}
        L_{Eu}=\frac{1}{MN}\\\sum_{i=1}^{M}\sum_{j=1}^{N}|| \hat{X}(i,j) - X(i,j) ||_2^2 
    \label{eucloss}
    \end{equation}

Here, $X$ and $\hat{X}$ denotes the $ M \times N$ ground truth or the noise free and despeckled or predicted images respectively.

In order to exploit the coherence property of pixels, we have defined the neighbourhood loss in eqn. (\ref{totalloss}) as, 
\begin{equation}
    \begin{split}
        L_N = \sqrt{\sum_{i=1}^{M}\sum_{j=1}^{N}{\hat{X}(i+1,j)} - \hat{X}(i,j))^2} \\+ \sqrt{\sum_{i=1}^{M}\sum_{j=1}^{N}{\hat{X}(i,j+1) - \hat{X}(i,j))^2}} \\+ \sqrt{\sum_{i=1}^{M}\sum_{j=1}^{N}{\hat{X}(i+1,j+1)} - \hat{X}(i,j))^2} \\+ \sqrt{\sum_{i=1}^{M}\sum_{j=1}^{N}{\hat{X}(i+1,j) - \hat{X}(i,j+1))^2}}
    \label{neighloss}
    \end{split}
\end{equation}

\begin{figure}[ht!]
\centering
    \includegraphics[scale = 0.3]{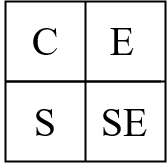}
    \caption{$2 \times 2$ Neighborhood}
    \label{fig:neigh}
\end{figure}

Fig. \ref{fig:neigh} shows a $2 \times 2$ neighborhood, where $C$, $E$, $S$ and $SE$ refers to the pixels at $(i,j)$, $(i,j+1)$, $(i+1,j)$ and $(i+1,j+1)$ respectively. The proposed neighbourhood loss is derived from the total variation (TV) loss \cite{idcnn}, which considers the horizontal, $C-E$ and vertical, $C-S$ differences only. In the proposed neighbourhood loss is designed to include the forward diagonal, $C-SE$ and reverse diagonal, $E-S$ differences in addition to those included in the TV loss. Thus, the neighborhood loss function attains the ability capture the coherence across all the neighbourhood pixels.

The perceptual loss could capture high-level semantic details of an image like edges and texture better than the per-pixel loss function \cite{perceptual}. Inspired by this idea, we have proposed a weighted perceptual loss function given by,

\begin{equation}
    L_{Per} = \frac{1}{ \rho MN}\\\sum_{k=0}^{n}\sum_{i=1}^{M}\sum_{j=1}^{N}{2^k(V_k(\hat{X}(i,j)) - V_k(X(i,j)))^2}
    \label{perloss}
\end{equation}

Here, $\rho = (2^{n+1}-1)$ and $V_k(:)$ denotes the output of the $k^{th}$ loss block of VGG16, whose weights are pre-trained on ImageNet dataset. In eqn. (\ref{perloss}), $V_0$ denotes the input to VGG16, and the output from the later layers uses the weighing term $2^k$ to gain more importance. We observed that feature extraction beyond $3$ VGG blocks leads to over-fitting, hence the no. of blocks, $n$ is set to 3. 
 
\section{Implementation Details}
\subsection{Training and Test Dataset}\label{dataset}
\begin{figure}[hb!]
    \centering 
\subfigure{
  \includegraphics[width=0.175\linewidth]{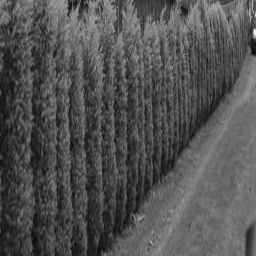}}\hspace{-1em}
\subfigure{
  \includegraphics[width=0.175\linewidth]{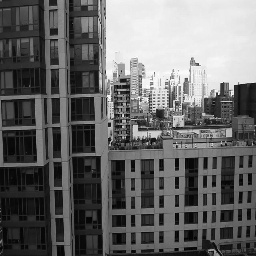}}\hspace{-1em}
\subfigure{
  \includegraphics[width=0.175\linewidth]{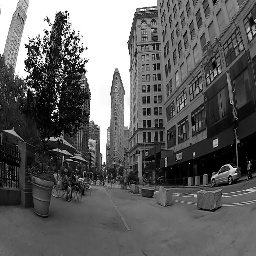}}\hspace{-1em}
\subfigure{
  \includegraphics[width=0.175\linewidth]{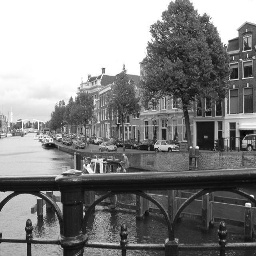}}\hspace{-1em}
\subfigure{
  \includegraphics[width=0.175\linewidth]{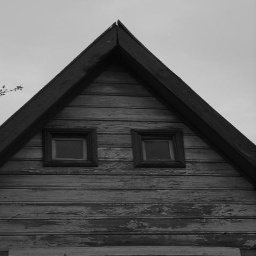}}\hspace{-1em}
\caption{Sample Images from the Dataset}
\label{sample}
\end{figure}

      We have generated the synthetic speckled dataset by mixing multiplicative Gamma noise and gray-scale images using eqn. (\ref{sar}). The dataset comprises of 3479 image pairs of size \textbf($256 \times 256$) obtained by varying the no. of looks, $L$ in eqn. (\ref{gamma}) from $2$ to $8$ in unit intervals and from $10$ to $30$ in increments of $5$. $2519$ image pairs out of the $3479$, are used for training, i. e. $229$ pairs for each of the 11 noise levels. Every look, $L$, is tested using 80 pairs of images. We have used a common dataset for training, and testing, to have a fair comparison between the benchmarking algorithms and the proposed method. Fig. \ref{sample}, portrays few sample images from the dataset. Two real SAR images are used to validate the despeckling performance of the proposed approach qualitatively. The Terrasar-X of Toulouse \copyright DLR is a X-band image with $L=1$ and $1.10m \times 1.04m$ resolution. The ERS-1 of Lelystadt is a band FIXME image with $L=3$ and has FIXME resolution. A sub-image of size $256 \times 256$, is considered for qualitative visual evaluation.
      
\subsection{Model Parameters}      
\begin{figure}[t!]
    \includegraphics[width=\linewidth]{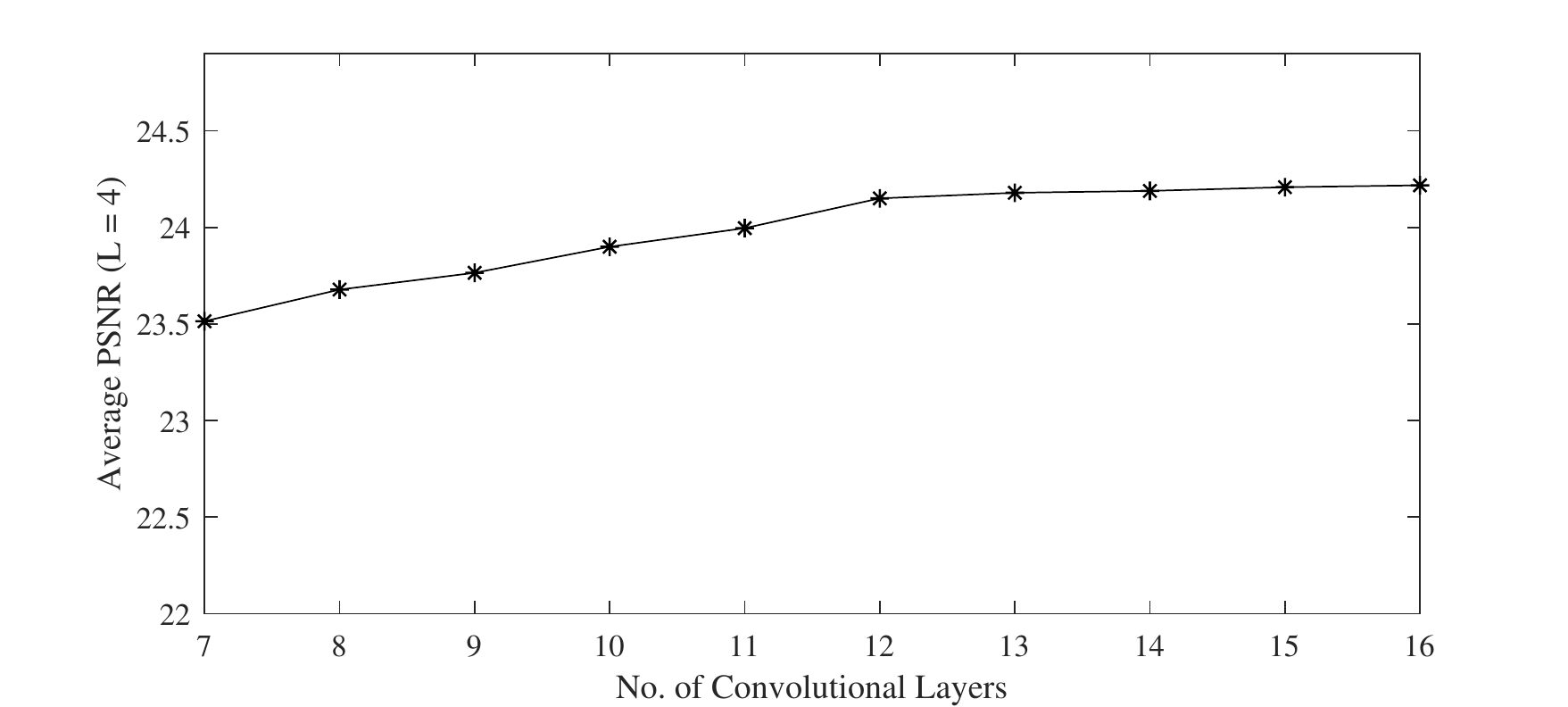}
    \caption{Average PSNR vs the Number of convolutional layers.}
    \label{fig:layers}
\end{figure}

The optimal value of the normalization parameter \(\alpha_N\) in eqn. (\ref{totalloss}) is obtained by varying it from $0.005$ to $0.00005$, and is found to be $0.0001$. Likewise, the optimal \(\beta_N\) is found to be $0.001$, after varying it between $0.002$ and $0.0005$. We have determined the optimal learning rate for Adam optimizer and the batch size at $0.0001$ and $16$. We trained the network until we observe a marginal change in the validation loss. An experiment to determine the optimal no. of convolutional layers for the proposed NeighCNN, is conducted. The no. of convolutional layers is varied between $7$ and $16$, and the average peak signal-to-noise ratio (PSNR) is determined for synthetic SAR images with $L = 4$. From fig. \ref{fig:layers}, it can be verified that the change in average PSNR is almost constant beyond 12 convolutional layers. From the literature, it can be verified that popular methods have fixed the no. of convolution layers based on intuition and have not validated it experimentally \cite{idcnn}-\cite{sardrn}. 

\begin{table*}[t]
\centering
\caption{Ablation Study on Loss Function}
\begin{tabular}{ccccccccc}
\hline
\hline
Look, L             & Metric & Noisy & L\_Per & L\_Eu & L\_Per + L\_N & L\_Eu + L\_N & L\_Eu + L\_Per & L\_Eu + L\_Per + L\_N \\ \hline
\multirow{3}{*}{1}  & PSNR 	 & 10.44 & 14.58  & 18.91 & 16.05         & 18.85         & 18.59         & \textbf{18.92} \\ 
                    & SSIM   & 0.17  & 0.27   & 0.48  & 0.46          & \textbf{0.49} & \textbf{0.49} & \textbf{0.49}  \\ 
                    & UQI    & 0.16  & 0.19   & 0.28  & 0.22          & \textbf{0.29} & 0.28          & \textbf{0.29}  \\ \hline
\multirow{3}{*}{2}  & PSNR   & 12.41 & 17.26  & 22.07 & 18.56         & 22.21        & 21.84          & \textbf{22.31}        \\  
                    & SSIM   & 0.24  & 0.38   & 0.58  & 0.54          & 0.59         & \textbf{0.60}  & \textbf{0.60}         \\  
                    & UQI    & 0.23  & 0.27   & 0.35  & 0.29          & 0.36         & 0.36           & \textbf{0.37}         \\ \hline
\multirow{3}{*}{5}  & PSNR   & 15.43 & 20.63  & 24.26 & 21.34         & 24.43        & 24.11          & \textbf{24.61}        \\  
                    & SSIM   & 0.35  & 0.54   & 0.69  & 0.65          & 0.70         & 0.71           & \textbf{0.72}         \\  
                    & UQI    & 0.34  & 0.38   & 0.45  & 0.38          & 0.46         & 0.45           & \textbf{0.47}         \\ \hline
\multirow{3}{*}{10} & PSNR   & 17.90 & 22.98  & 25.79 & 22.97         & 26.01        & 25.59          & \textbf{26.25}        \\  
                    & SSIM   & 0.44  & 0.65   & 0.76  & 0.71          & 0.77         & 0.77           & \textbf{0.78}         \\  
                    & UQI    & 0.43  & 0.46   & 0.52  & 0.45          & 0.53         & 0.52           & \textbf{0.54}         \\ \hline
\multirow{3}{*}{15} & PSNR   & 19.37 & 24.23  & 26.60 & 23.74         & 26.84        & 26.39          & \textbf{27.12}        \\  
                    & SSIM   & 0.49  & 0.70   & 0.79  & 0.73          & 0.80         & 0.80           & \textbf{0.81}         \\  
                    & UQI    & 0.48  & 0.49   & 0.55  & 0.48          & 0.56         & 0.55           & \textbf{0.57}         \\ \hline
\multirow{3}{*}{20} & PSNR   & 20.46 & 25.04  & 27.20 & 24.22         & 27.43        & 26.99          & \textbf{27.77}        \\  
                    & SSIM   & 0.53  & 0.73   & 0.81  & 0.75          & 0.82         & 0.81           & \textbf{0.83}         \\  
                    & UQI    & 0.51  & 0.52   & 0.57  & 0.50          & 0.58         & 0.57           & \textbf{0.59}         \\ \hline
\hline
\end{tabular}
\label{tab:abla}
\end{table*}

\begin{table*}[t!]
\centering
\caption{Quantitative Analysis of NeighCNN}
\begin{tabular}{cccccccc}
\hline
\hline
Look, $L$            & Metric & Noisy & Kuan  & SAR-BM3D & SARDRN & IDCNN & NeighCNN       \\ \hline
\multirow{3}{*}{1}  & PSNR   & 10.44 & 14.00 & 16.07    & 17.87  & 16.48 & \textbf{18.92} \\  
                    & SSIM   & 0.17  & 0.30  & 0.47     & 0.44   & 0.35  & \textbf{0.49}  \\ 
                    & UQI    & 0.16  & 0.24  & 0.24     & 0.28   & 0.24  & \textbf{0.29}  \\ \hline
\multirow{3}{*}{2}  & PSNR   & 12.41 & 16.15 & 17.81    & 20.96  & 21.09 & \textbf{22.31} \\  
                    & SSIM   & 0.24  & 0.36  & 0.51     & 0.54   & 0.55  & \textbf{0.60}  \\  
                    & UQI    & 0.23  & 0.29  & 0.31     & 0.33   & 0.33  & \textbf{0.37}  \\ \hline
\multirow{3}{*}{5}  & PSNR   & 15.43 & 18.72 & 19.49    & 23.25  & 23.93 & \textbf{24.61} \\  
                    & SSIM   & 0.35  & 0.45  & 0.57     & 0.65   & 0.69  & \textbf{0.72}  \\  
                    & UQI    & 0.34  & 0.37  & 0.41     & 0.42   & 0.43  & \textbf{0.47}  \\ \hline
\multirow{3}{*}{10} & PSNR   & 17.90 & 20.26 & 20.99    & 24.41  & 25.33 & \textbf{26.25} \\  
                    & SSIM   & 0.44  & 0.52  & 0.63     & 0.71   & 0.74  & \textbf{0.78}  \\  
                    & UQI    & 0.43  & 0.41  & 0.48     & 0.47   & 0.49  & \textbf{0.54}  \\ \hline
\multirow{3}{*}{15} & PSNR   & 19.37 & 20.94 & 21.96    & 24.94  & 25.99 & \textbf{27.12} \\  
                    & SSIM   & 0.49  & 0.56  & 0.66     & 0.74   & 0.76  & \textbf{0.81}  \\  
                    & UQI    & 0.48  & 0.43  & 0.52     & 0.49   & 0.51  & \textbf{0.57}  \\ \hline
\multirow{3}{*}{20} & PSNR   & 20.46 & 21.35 & 22.78    & 25.21  & 26.42 & \textbf{27.77} \\  
                    & SSIM   & 0.53  & 0.58  & 0.69     & 0.75   & 0.78  & \textbf{0.83}  \\  
                    & UQI    & 0.51  & 0.44  & 0.54     & 0.50   & 0.53  & \textbf{0.59}  \\ \hline
\hline
\end{tabular}
\label{tab:psnr}
\end{table*}   

\section{Experimental Results}
The proposed NeighCNN algorithm is tested initially on simulated SAR images, in terms of peak signal to noise ratio (PSNR), structural similarity index (SSIM), and universal image quality index (UQI). PSNR, SSIM, UQI metrics can only be computed on synthetic SAR images, as they require noise-free images as ground truth. PSNR quantifies the ability of the despeckling algorithm to smooth noise whereas, SSIM confirms its ability to preserve structural information like edges. UQI is a universal metric that does not depend on the image as well as its viewing conditions. The despeckling performance of NeighCNN is then verified qualitatively on real SAR images.
   
\subsection{Ablation Study on Loss Function} 
We have conducted an ablation study to validate the impact of the proposed three-part loss function. Several combinations of the loss function, Euclidean, perceptual, and neighbourhood are considered for training the model. In table \ref{tab:abla}, we have summarized the PSNR, SSIM, and UQI metrics for the different loss functions used in this study. It can be verified that some loss functions work well for specific metric and fail to outperform in terms of the other. However, the proposed three-part loss function is superior to the remaining combinations in terms of PSNR, SSIM as well as UQI for $L = 1, 2, 5, 10, 15$ and $20$. 

\subsection{Performance Analysis}
In this section, the performance of the proposed algorithm is verified quantitatively on synthetic and qualitatively on real SAR images. Popular benchmarking algorithms like, Kuan spatial domain filter, SAR block matching algorithm, SARBM3D \cite{sarbm3d}, IDCNN \cite{idcnn}, and SARDRN \cite{sardrn}, are considered for performance analysis. We have used the MatLab executable of the SARBM3D algorithm for experimentation \cite{grip}. We have set the hyper-parameters of the benchmarking algorithms, as mentioned in their respective papers.

Average results of the 80 test images are tabulated in table \ref{tab:psnr} for $L = 1, 2, 5, 10, 15$, and $20$. SARDRN outperforms IDCNN for the highest noise level at $L = 1$. For all the other noise levels, IDCNN is superior to all the other benchmarking algorithms. NeighCNN demonstrates an increase of $0.68 - 2.44$dB, $0.03 - 0.14$, and $0.04 - 0.06$ in terms of PSNR, SSIM, and UQI when compared to IDCNN for the listed noise levels. 

\begin{figure*}[t!]
    \centering 
\subfigure{
  \includegraphics[width=0.14\linewidth]{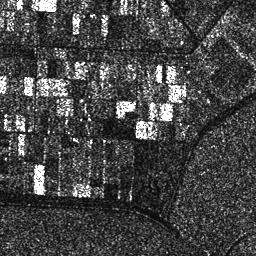}
  \label{ln}}
\subfigure{
  \includegraphics[width=0.14\linewidth]{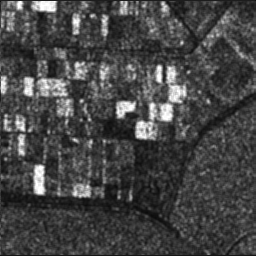}
  \label{lk}}
\subfigure{
  \includegraphics[width=0.14\linewidth]{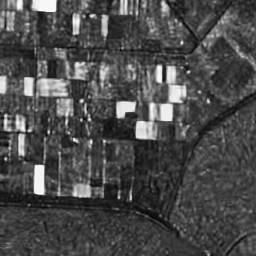}
  \label{lb}}
\subfigure{
  \includegraphics[width=0.14\linewidth]{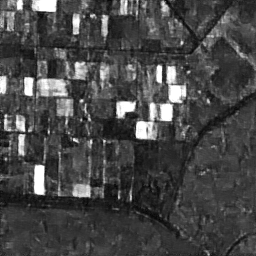}
  \label{ld}}
\subfigure{
  \includegraphics[width=0.14\linewidth]{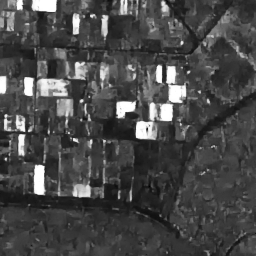}
  \label{lid}}
\subfigure{
  \includegraphics[width=0.14\linewidth]{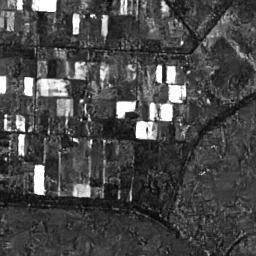}
  \label{lo}}

\medskip
\subfigure{
  \includegraphics[width=0.14\linewidth]{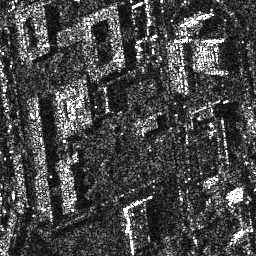}
  \label{tn}}
\subfigure{
  \includegraphics[width=0.14\linewidth]{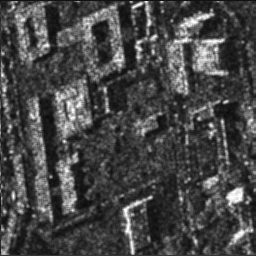}
  \label{tk}}
\subfigure{
  \includegraphics[width=0.14\linewidth]{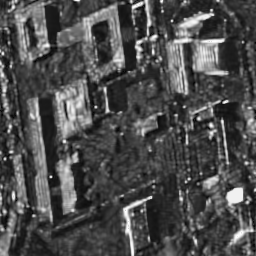}
  \label{tb}}
\subfigure{
  \includegraphics[width=0.14\linewidth]{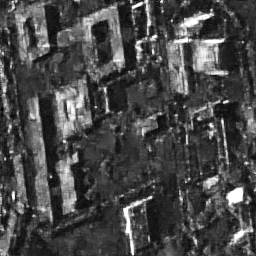}
  \label{td}}
\subfigure{
  \includegraphics[width=0.14\linewidth]{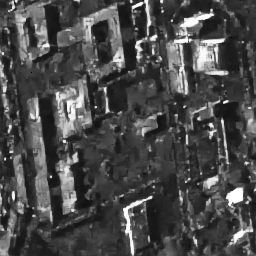}
  \label{tid}}
\subfigure{
  \includegraphics[width=0.14\linewidth]{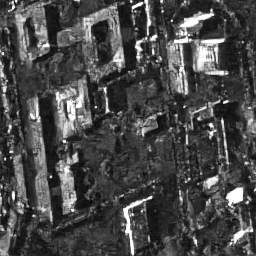}
  \label{to}}
\caption{From left to right: Terrasar-X image of Toulouse \copyright DLR (Top), ERS-1 image of Lelystadt \copyright ESA (Bottom), Kuan Filter, SAR-BM3D, SARDRN, IDCNN and NeighCNN.}
\label{visComp}
\end{figure*}

For visual evaluation of the proposed approach, two real SAR test images, are considered. The first column in fig. \ref{visComp} shows the original SAR images.  Subsequent columns in fig. \ref{visComp} portrays the despeckled images using Kuan spatial filter, SARBM3D, SARDRN, IDCNN, and NeighCNN algorithms.  We can verify that the proposed NeighCNN architecture achieves improved noise removal and edge preservation capabilities. Traditional measure for evaluating the performance of the despeckling algorithm, equivalent no. of looks (ENL), is not demonstrated on the real SAR images as we believe that ENL quantifies the smoothing capabilities but not the edge preservation ability. After several experiments, we observed that over-smoothed images yield higher values of ENL.

\section{Conclusion}
    In this work, we have proposed a CNN based despeckling approach for SAR images. Based on an extensive experiment, the optimal number of convolutional layers for the proposed architecture is determined. NeighCNN architecture can effectively capture the non-linear relationship between the speckled image and its ground truth. Experimental results verify that the subtraction layer is sufficient to handle the multiplicative nature of the noise instead of complex operations like logarithm-exponential pair with subtraction layer \cite{sarcnn} or tanh followed by a division layer \cite{idcnn}. The PSNR, SSIM, and UQI measures of the unique total loss function validate the capability of NeighCNN to minimize both the low-level as well as high-level differences between the ground truth and the despeckled image. Thus, the proposed architecture achieves an optimal trade-off between speckle reduction and feature preservation. In the future, generative adversarial networks (GANs) can replace CNNs to perform blind despeckling in SAR images.

\bibliographystyle{IEEEtran}

\end{document}